\documentclass[letterpaper]{JHEP3}
\usepackage{cite}

\newcommand{\be}{\begin{equation}}
\newcommand{\ee}{\end{equation}}
\newcommand{\bea}{\begin{eqnarray}}
\newcommand{\eea}{\end{eqnarray}}
\newcommand{\N}{\mathcal{N}}

\renewcommand{\t}[1]{\tilde{#1}}
\newcommand{\Del}{\nabla}
\newcommand{\del}{\partial}
\newcommand{\bi}{\bar{\imath}}
\newcommand{\bj}{\bar{\jmath}}
\newcommand{\comment}[1]{}

\title{Notes on $SU(3)$ Structures in Type IIB Supergravity}
\author{Andrew R. Frey\\ California Institute of Technology\\ 
Mail Stop 452-48\\ Pasadena, CA 91125, USA\\
\email{frey@theory.caltech.edu}}
\preprint{hep-th/0404107 \\ CALT-68-2492}
\abstract{We study
solutions of type IIB
supergravity with an $SU(3)$ structure group and four dimensional Poincar\'e
invariance and present relations among the bosonic fields 
which follow from the supersymmetry variations.  We make explicit some 
results which also follow from the more general case of an $SU(2)$ 
structure and give some short comments applicable to general supersymmetric
solutions.  We also provide simplified relations appropriate for duals of
gauge theory renormalization group flows, and use these to 
derive the supergravity solution for a bound state of $(p,q)$5-branes
and D3-branes.}

\begin{document}
\section{Introduction}\label{s:intro}
Starting with the analysis of heterotic string 
compactifications \cite{Candelas:1985en,Strominger:1986uh}, it has been
known that the conditions for unbroken supersymmetry, that is, the vanishing
of the supersymmetry variations of the supergravity fermions, are often much
easier to solve than the equations of motion.  Indeed, there are many 
recent examples in which the supersymmetry variations have been used to find
interesting supergravity duals of gauge theories\footnote{There is a very
large literature of examples, and it is beyond the scope of this paper to 
list them all.  We will cite some specific examples as they come up.}.  
Therefore, and also because classically supersymmetric
compactifications are important candidates for the vacuum of string theory,
a systematic analysis of the supersymmetry conditions is of great intrinsic
interest.  

Such an analysis has been carried out for M-theory, 11 dimensional
supergravity, (and by extension type IIA 10 dimensional supergravity) over
the past decade, beginning with 
\cite{Becker:1996gj,Becker:2000jc,Becker:2000rz}.  Very great progress has
been made recently by understanding the importance of the structure group in
supersymmetry; analysis of M-theory solutions can be found in
\cite{Gauntlett:2002fz,Kaste:2003dh,Behrndt:2003uq,Kaste:2003zd,
Martelli:2003ki,Behrndt:2003zg,Dall'Agata:2003ir,Gauntlett:2003wb,
Behrndt:2003ih,Gauntlett:2004zh,Becker:2004qh,Behrndt:2004km}.

The situation in type IIB supergravity has until recently 
focused much more on specific solutions and particularly simple classes of
solutions.  Of great importance have been the type B solutions, which 
have a warped Calabi-Yau geometry, imaginary self-dual and primitive 3-forms,
and a 5-form determined by the warp factor; the supersymmetry conditions
in these solutions were studied in 
\cite{Kehagias:1998gn,Grana:2000jj,Gubser:2000vg,Grana:2001xn}.  
Compactified type B solutions have been studied extensively; see
\cite{Frey:2003tf} for a review.
There is also a class, type A, of solutions dual to the heterotic solutions of
\cite{Strominger:1986uh}, which have been analyzed in the group structure
language for all supergravities in \cite{Gauntlett:2003cy}.
(Some compactified type A (and S-dual) solutions in type IIB string theory
are discussed in \cite{Becker:2002sx,Kachru:2002sk}; we do not discuss
the widely studied heterotic compactifications here.)
Then \cite{Frey:2003sd} discussed a more general class of solutions which
interpolate between the type B class and the S-dual of the type A solutions,
which include D5-brane solutions.  Recently\footnote{In fact, this paper was
in preparation when the results of \cite{Dall'Agata:2004dk} appeared.}, 
\cite{Dall'Agata:2004dk}
presented an analysis of $SU(2)$ structures in type IIB supergravity; this
class of solutions includes type A and B solutions as well as particular
solutions corresponding to polarized 5-branes 
\cite{Pilch:2003jg,Pilch:2004yg}.  
This paper will report
the conditions required for $SU(3)$ structures in IIB supergravity, assuming
a four dimensional Poincar\'e invariance.  Although $SU(3)$ structures are
a subset of $SU(2)$ structures, the constraints relating different fields
degenerate for a pure $SU(3)$ structure, so it is useful to have a direct
analysis (the precise nature of the degeneration is explained in 
\cite{Frey:2003sd} and will be discussed in section \ref{s:radial}).   
(Note that $SU(3)$ structures in type IIB supergravity have appeared before
in studies of mirror symmetry \cite{Fidanza:2003zi}.)

We begin in section \ref{s:constraints} by giving the supersymmetry 
variations in string frame and then converting them to relations among
the bosonic fields and $SU(3)$ structure.  We present those relations in
\ref{ss:relate} after showing in section \ref{ss:nijenhuis} that the
complex structure is integrable, and we make a few general comments in
\ref{ss:other}.  In section \ref{s:radial}, we integrate the scalar
relations in the case that all scalars depend only on one coordinate, 
such as a radius.  These results could be helpful for finding supergravity
duals of renormalization group flows.  Finally, in section \ref{s:bound},
we use our results to derive the supergravity solution for a bound state
of a $(p,q)$5-brane and D3-branes.

\section{Constraints from $SU(3)$ Structure}\label{s:constraints}

The fermion supersymmetry variations for ten dimensional IIB supergravity
are given in \cite{Schwarz:1983wa,Schwarz:1983qr}.  We use the modern string
theory conventions, in which the variations are 
\cite{Bergshoeff:1999bx,Hassan:1999bv}
\bea
\delta\lambda &=& \frac{1}{2}\left(\del_M\phi-i e^\phi \del_M C
\right)\Gamma^M\varepsilon +\frac{1}{24}\left(ie^\phi \t F-H\right)_{MNP}
\Gamma^{MNP}\varepsilon^*    \label{dilvar}\\
\delta\psi_M &=&  \left( \Del_M+\frac{i}{8}e^\phi\del_N C\, \Gamma^N\Gamma_M
+\frac{i}{16\cdot 5!} e^\phi \t F_{NPQRS}\Gamma^{NPQRS}\Gamma_M\right)
\varepsilon\nonumber\\
&&-\left(\frac{1}{8}H_{MNP}\Gamma^{NP}+\frac{i}{48}e^\phi\t F_{NPQ}\Gamma^{NPQ}
\Gamma_M\right)\varepsilon^*\ .    \label{gravvar}
\eea
The supersymmetry parameter $\varepsilon$ is a 10D Weyl spinor of positive
chirality; the rest of the conventions are explained in \ref{a:conventions}.

Since we are working in a vacuum with 4D Poincar\'e invariance, the most
general ansatz we can take is
\bea
ds^2 &=& e^{2A}\eta_{\mu\nu}dx^\mu dx^\nu + g_{mn}dx^m dx^n\nonumber \\
\t F_{\mu\nu\lambda\rho m} &=& e^{-4A}\epsilon_{\mu\nu\lambda\rho}\del_m h
\ .\label{ansatz}\eea
All other form components and all field dependences are in the internal 
directions.  We include the factor $e^{-4A}$ in the 5-form so that it
is an external derivative in those components.  
Field strengths without tildes are just external
derivatives while tildes denote the addition of Chern-Simons terms.  The only
definition of importance to us is $\t F_3 = F_3-C H_3$, as the 5-form is 
given through its self-duality.

\subsection{Normalization}\label{ss:norm}

In this subsection and the following one, we will emphasize two points 
which were not stated explicitly in \cite{Dall'Agata:2004dk}.  The first
point regards the normalization of the spinors.  If we start by taking
\be\label{susy1}
\varepsilon = e^{A/2}\left(\zeta\otimes \chi_+ +\zeta^*\otimes \chi_-\right)
\ ,\ee
where $\zeta$ is a 4D Weyl spinor of positive chirality and $\chi_\pm$ are
6D Weyl spinors of respective chirality, the internal and external gravitino
variations (\ref{gravvar}) combine to give
\bea
\left[\Del_m+\frac{i}{4}e^\phi\del_m C -\frac{1}{2}\del_n A\gamma_m{}^n
+\frac{1}{4}e^{\phi-4A}\del_n h\gamma_m{}^n\right]\chi_+-\frac{1}{8}
\left(H+ie^\phi \t F\right)_{mnp}\gamma^{np}\chi_-^* &=&0\nonumber\\
\left[\Del_m+\frac{i}{4}e^\phi\del_m C -\frac{1}{2}\del_n A\gamma_m{}^n
-\frac{1}{4}e^{\phi-4A}\del_n h\gamma_m{}^n\right]\chi_- -\frac{1}{8}
\left(H+ie^\phi \t F\right)_{mnp}\gamma^{np}\chi_+^* &=&0\! .\label{normvar}
\eea

We can immediately see that
\bea
\Del_m\left(\chi_+^\dagger\chi_+ +\chi_-^\dagger\chi_-\right) &=&
\frac{1}{8}\left(H+ie^\phi\t F\right)_{mnp}
\left(\chi_+^\dagger\gamma^{np}\chi_-^* +\chi_-^\dagger\gamma^{np}\chi_+^*
\right)\nonumber\\
&&-\frac{1}{8}\left(H-ie^\phi\t F\right)_{mnp}\left(
\chi_-^T \gamma^{np}\chi_+ + \chi_+^T\gamma^{np}\chi_-\right)\ .
\label{normderiv}\eea
This vanishes by the symmetry properties of the $\gamma^m$ matrices.
We easily see, therefore, that the $SU(2)$ structure spinors of 
\cite{Dall'Agata:2004dk} (in the notation of that paper) satisfy 
\be\label{su2norm}
e^{-A}\left(|a|^2+|b|^2+|c|^2\right)=\textnormal{constant}\equiv 1\ .\ee
(The warp factor must be included because \cite{Dall'Agata:2004dk} does
not scale the spinor by an overal power of the warp factor.)

In an $SU(3)$ structure, there is only one linearly independent 6D spinor,
which corresponds to $c\to 0$ in the $SU(2)$ structure notation of
\cite{Dall'Agata:2004dk}.  We can therefore write our supersymmetry
parameters as
\be\label{su3spinor}
\varepsilon = e^{A/2}\left( \cos(\alpha/2)e^{i\beta/2}\zeta\otimes \chi
+\sin(\alpha/2) e^{i\beta/2}\zeta^*\otimes\chi^*\right)\ee
with $\chi$ a positive chirality 6D spinor with $\chi^\dagger\chi=1$.

A number of recent papers 
\cite{Gowdigere:2003jf,Pilch:2003jg,Nemeschansky:2004yh,Pilch:2004yg}
have used an ``algebraic spinor'' approach to derive supergravity duals to
renormalization group flows.  In this approach, the spinors are specified by
projector equations; the projector corresponding to (\ref{su3spinor})
(that is, such that $P\varepsilon=\varepsilon$) is 
\be\label{su3projector}
P=\frac{1}{2}\left[ 1+\cos\alpha \gamma_{(\hat 6)}+\sin\alpha e^{i\beta} *
\right]\ ,\ee
where $\gamma_{(\hat 6)}$ is the six dimensional chirality (alternately, the 
four dimensional chirality can be used) and $*$ is the complex conjugation
operator.  Additional projectors are necessary to reduce the number of 
supersymmetries to $\N=1$; these simply align the spins of $\chi$ (see the
final paragraph of \ref{ss:nijenhuis}).

\subsection{Complex Structure Integrability}\label{ss:nijenhuis}
The other point which we wish to make explicit concerns the integrability of
the almost complex structure (ACS). 
Any $SU(3)$ structure has an almost complex
structure tensor as well as a $(3,0)$ form defined with respect to that
ACS; these forms give an alternate definition of the $SU(3)$ structure in
6D.  Defining the ACS 
\be\label{acs}
J_m{}^n= -i\chi^\dagger\gamma_m{}^n\chi\ ,\ee
Fierz identities (see appendix \ref{a:conventions}) show that it indeed squares
to $-1$.  See, for example, \cite{Gauntlett:2003cy} for a nice review of
different group structures in different dimensions.

The ACS is integrable if the Nijenhuis tensor 
$N_{mn}{}^p = J_m{}^q\Del_{[q}J_{n]}{}^p-J_n{}^q\Del_{[q}J_{m]}{}^p$
vanishes.  
From (\ref{normvar}), we find
\bea
\Del_m J_n{}^p &=& -\left(\del_q A -\frac{1}{2}\cos\alpha e^{\phi-4A}\del_q
h \right)\left(g_{mn}J^{pq}-\delta_m^pJ_n{}^q+\delta_n{}^q J_m{}^p-g^{pq}
J_{mn}\right)\nonumber\\
&& +\frac{1}{2}T_{mn}{}^rJ_r{}^p -\frac{1}{2}T_{mr}{}^pJ_n{}^r\ ,
\label{jderiv}\\
T_{mnp}&\equiv&\sin\alpha\left( \cos\beta H+\sin\beta e^\phi \t F\right)_{mnp} 
\ .\label{torsion}\eea
The flux tensor $T$ acts like a torsion in the connection, and the other terms
in $\Del J$ are similar to those generated in a connection by rescaling the
metric.  In fact, it is a short calculation to see that the terms with $A$
and $h$ contribute nothing to the Nijenhuis tensor; the only contribution is
from the torsion.

However, one linear combination of equations that result from the dilatino
variation (\ref{dilvar}) is 
\be\label{diltorsion}
\frac{1}{12}T_{mnp}\gamma^{mnp}\chi = \left[\left(\cos(2\beta)+i\cos\alpha
\sin(2\beta)\right)\del_m\phi+\left(\sin(2\beta)-i\cos\alpha\cos(2\beta)\right)
e^\phi\del_m C\right]\gamma^m\chi\ .\ee
Once we have equation (\ref{diltorsion}), \cite{Strominger:1986uh} shows
that the torsion contributes nothing to the Nijenhuis tensor.  Therefore,
for an $SU(3)$ structure, the complex structure is always integrable!  In fact,
this follows from equation (3.33) in \cite{Dall'Agata:2004dk}; this is a
more explicit statement and proof.  (This proof also ties up a loose end
in \cite{Frey:2003sd}.)

With an integrable complex structure, there is a holomorphic atlas, so we 
can write the defining relation for a Hermitean metric $J_{i\bj}=ig_{i\bj}$
in the complex coordinates.  This, along with (\ref{acs}), implies that
$\gamma_{i}\chi=\gamma^{\bi}\chi=0$.

\subsection{Relations Among Supergravity Fields and Geometry}\label{ss:relate}

The vanishing of the fermion variations (\ref{dilvar},\ref{gravvar}) implies
a number of constraints on the supergravity fields and the $SU(3)$ structure
of the manifold.  Here, we briefly report those constraints, which are a
special case of those reported in \cite{Dall'Agata:2004dk} for $SU(2)$ 
structures.  We write these constraints directly in terms of the supergravity
fields, rather than decomposing the fields according to the $SU(3)$ 
structure.

There are four independent equations relating the scalars $\alpha$, $\beta$,
$A$, $h$, $\phi$, and $C$.  These relations, in the complex coordinates
defined by $J$, are
\bea
b_1\del A +\frac{1}{4} a_1\left( ie^\phi\del C-e^{\phi-4A}\del h\right) &=& 0
\nonumber\\
a_1\del A -\frac{1}{2}a_1\del\phi -\frac{1}{4}b_1\left(ie^\phi\del C +
e^{\phi-4A}\del h\right)&=&0\nonumber\\
\frac{1}{2}\sin(2\alpha)\del\alpha -\cos\alpha\left(a_2\del\phi +ib_2
e^\phi\del C\right) &=&0\nonumber\\
\sin^2\alpha \left(\del\beta+\frac{1}{2}e^\phi\del C+\frac{i}{2}e^{\phi-4A}
\del h\right) + \cos\alpha\left(ib_2\del\phi-a_2e^\phi\del C\right)&=&0
\label{scalars} 
\eea
where
\bea
a_1=\cos\alpha\cos\beta-i\sin\beta&,&b_1=\cos\beta-i\cos\alpha\sin\beta
\nonumber\\
a_2=\cos\alpha\cos(2\beta)-i\sin(2\beta)&,&b_1=\cos(2\beta)
-i\cos\alpha\sin(2\beta)
\ .\label{ab}\eea
In the third equation of (\ref{scalars}), we have left in the overall factor
of $\cos\alpha$ to demonstrate how the relations can degenerate at loci in
the space of spinors.  Finally, we note that it is simple to translate these
equations to real coordinates by using the complex structure.  We simply take
the holomorphic derivative to the exterior derivative $\del f\to df$ when
no factors of $i$ are present and similarly
$i\del f\to J_m{}^n\del_n f dx^n$.  The integrability conditions for 
(\ref{scalars}) are quite complicated, and we have not done a full analysis.
However, in certain cases, such as $\alpha\to 0,\pi/2$ or $\beta\to 0,\pi/2$,
the integrability conditions (and scalar relations) simplify considerably.
The only new condition introduced in any of those cases is that 
$\del\phi\wedge \del C=0$.  In section \ref{s:radial}, we will discuss a
class of solutions in which we can integrate (\ref{scalars}) explicitly.

There are numerous algebraic and differential relations for the 3-forms and
$SU(3)$ structure.  We start by defining the $(3,0)$ form of the structure,
\be\label{omega}
\Omega_{mnp}=\chi^T\gamma_{mnp}\chi\ .\ee
Then it is straightforward to see 
\be\label{no30}
H_{mnp}\Omega^{mnp}=\t F_{mnp}\Omega^{mnp}=0\ee
and the conjugate equations, so there are no $(3,0)$ or $(0,3)$ components of
the fluxes.  There are also the following relations for the fluxes:
\bea
\left(\cos\alpha\cos\beta\delta_m^n-\sin\beta J_m{}^n\right)\del_n\phi
+\left(\cos\alpha\sin\beta \delta_m^n+\cos\beta J_m{}^n\right)e^\phi\del_n C
&=& \frac{1}{4}\sin\alpha e^\phi \t F_{mnp}J^{np}\ \ \ \ \ \ \ 
\ \ \label{Hsingle}\\
\left(\cos\alpha\sin\beta \delta_m^n+\cos\beta J_m{}^n\right)\del_n\phi
-\left(\cos\alpha\cos\beta\delta_m^n-\sin\beta J_m{}^n\right)e^\phi \del_n C
&=&\frac{1}{4}\sin\alpha H_{mnp}J^{np}\ \ \ \ \ \ \ \ \ \label{Fsingle}
\eea
and
\bea
-\frac{1}{4}\left(\sin\beta H-\cos\beta e^\phi\t F
\right)_{mnp}J^{np}&=&\del_m\alpha \label{HFalpha}\\
-\frac{1}{4}\cos\alpha\left(\cos\beta
H+\sin\beta e^\phi \t F\right)_{mnp}J^{np}
&=&\sin\alpha\left(\del_m\beta+\frac{1}{2}e^\phi \del_m C+\frac{1}{2}
e^{\phi-4A}\del_n h J_m{}^n\right).\ \ \ \ \ \ \label{HFbeta}\eea
After a bit more work and application of equation (\ref{fierz1}),
\bea
0&=& \left[ 2\sin\alpha e^{\phi-4A}g_{m[n}\del_{p]}h+
\cos\alpha\left(\cos\beta H+\sin\beta e^\phi\t F\right)_{mnp}+\right.
\nonumber\\
&& \left. i\left(\sin\beta H-\cos\beta e^\phi\t F\right)_{mnp}\right]
\left(\delta_q^n+iJ_q{}^n\right) \left(\delta_r^p+iJ_r{}^p\right)\ .\ \ 
\label{12project}\eea
In the type B, $\alpha=0$, ansatz, this equation requires that $G=\t F
-ie^{-\phi}H$ is $(1,2)$ only.  This is opposite the usual convention for
choice of complex coordinates.

The other relations are differential in the complex structure and $(3,0)$
form (here $T$ is the torsion part of the 3-forms, as defined in equation
(\ref{torsion})):
\bea
dJ&=& \left(2dA-\cos\alpha e^{\phi-4A}dh+2\cos(2\beta)d\phi+2\sin(2\beta)
e^\phi dC\right)\wedge J\nonumber\\
&&+\star\left[T-2\cos\alpha\left(\sin(2\beta)d\phi-\cos(2\beta)e^\phi dC
\right)\wedge J\right]\ , \ \label{dJ}\\
\Del_n J_m{}^n&=& \left(4\del_n A-2\cos\alpha e^{\phi-4A}\del_n h+
2\cos(2\beta)\del_n\phi +2\sin(2\beta)e^\phi\del_n C \right) J_m{}^n
\nonumber\\
&&+2\cos\alpha\left(\sin(2\beta)\del_m\phi - \cos(2\beta)e^\phi\del_m C\right)
\ , \label{dstarJ}\\
d\Omega &=& \left[3dA-\frac{3}{2}\cos\alpha e^{\phi-4A}dh+i\cos\alpha
\left( d\beta+\frac{1}{2}e^\phi dC\right)\right.\nonumber\\
&&\left. +2\cos(2\beta)d\phi+2\sin(2\beta)e^\phi dC\right]\wedge \Omega
\ .\label{domega}
\eea
There is no separate equation for $d\star_6 \Omega$ because $\Omega$ satisfies
a self-duality relation in the internal space.  Note that it is often possible
to integrate the $dh$ terms, in which case the $dA,dh$ terms both can be
interpreted as due to rescaling the six dimensional metric.  

\subsection{Other Concerns}\label{ss:other}

The relations described in section \ref{ss:relate}, while containing 
information about the supersymmetry of a particular string vacuum, do not
suffice to specify the solution entirely.  It is also necessary to 
apply the Bianchi identities for the field strengths and to check that the
solution satisfies the equations of motion.  The necessary Bianchi 
identities are
\bea
dH=0&,& dF_3 =0 \textnormal{ or } d\t F_3 = -dC\wedge H\nonumber\\
d\t F_5 &=& H_3\wedge F_3\ .\label{bianchi}
\eea
Note that the 5-form Bianchi identity is also an equation of motion due to
the self-duality of the 5-form.

The reason we should also check the equations of motion is that not all the
equations of motion are necessarily 
contained in the commutators of the supersymmetry variations.  This issue
was discussed in \cite{Gauntlett:2002fz} for the case of M-theory.  
However, in IIB supergravity,
\cite{Schwarz:1983qr} was able to derive all the equations of motion by
using the supersymmetry algebra.  It seems, therefore, that because the 
5-form Bianchi is also an equation of motion, supersymmetry of a type IIB
background, along with the Bianchi identities, 
may imply that it solves the equations of motion.

\section{Radial Flow}\label{s:radial}

In this section, we integrate the scalar relations (\ref{scalars}) 
in a case that should be relevant for renormalization group flows in 
string-gauge dualities as well as for supersymmetric brane bound states.
Specifically, we consider the case that all the scalars depend only on
a single (real) dimension, such as a radial direction.  This ansatz could
be useful for finding supergravity solutions relating the supergravity 
backgrounds of \cite{Maldacena:2000yy} and \cite{Klebanov:2000hb}, 
for example, as both are $\N=1$ backgrounds with
$SU(3)$ structure (see \cite{Papadopoulos:2000gj} and 
\cite{Grana:2000jj,Gubser:2000vg} for supersymmetry analysis, respectively).

Once we specify that the scalars depend only on a single direction $r$,
we can separate (\ref{scalars}) into real and imaginary parts by factoring
out $\del r/\del z^i$.  The eight equations we have are more than enough 
to define all the scalars in terms of one of the angles $\alpha,\beta$,
and in fact the extra equations are compatible with the following solutions.
In determining the proportionality factors in the following, we assumed
that as $r\to\infty$, $\beta\to\theta$ and $\alpha\to\pi/2-\theta'$ with 
$\theta,\theta'\neq 0,\pi/2$.  We also took $\phi\to\ln g_s$, $C\to 0$, 
and $A\to 0$ as boundary conditions; these are appropriate for the 
asymptotically flat region around a brane bound state.  However, it is 
possible to adjust the proportionality constants to allow for other boundary
conditions, as would be the case in string-gauge dualities.  We have
\bea
\cos\alpha&=& \sin\theta'\cot\theta \tan\beta\nonumber\\
e^\phi &=& g_s\frac{\sin(2\theta)}{\sin(2\beta)}\nonumber\\
C&=&\frac{\tan\theta}{g_s}\left(1-\frac{\sin^2\beta}{\sin^2\theta}\right)
\nonumber\\
e^{2A}&=& \frac{\sin\theta\cot\theta'}{\sin\beta\tan\alpha}=
\frac{\cos\theta'\cos\theta}{\sin\alpha\cos\beta}\nonumber\\
h&=& \frac{1}{g_s}\cot^2\theta' \sin\theta' \cot^2\alpha +h_0\label{flow}
\eea
with $h_0$ some constant (which we can fix to zero by a gauge transformation).

This is an appropriate point to emphasize how the scalar relations 
(\ref{scalars}) degenerate in certain limits, which makes the integrated
scalars (\ref{flow}) singular.  The ``renormalization'' necessary to 
interpret (\ref{flow}) in such a limit is discussed in detail in
\cite{Frey:2003sd}, where it was argued that the proportionality constants
(ie, the angles $\theta,\theta'$) can absorb zeros and infinities.
As an example, suppose that $\beta(x)=\pi/2 -\epsilon z(x)$, where 
$\epsilon\to 0$ in the limit we consider.  To renormalize (\ref{flow}), then
let also $\theta=\pi/2-\epsilon$.   By being careful, we get the new
relations
\bea
\cos\alpha&=& \sin\theta' z^{-1}\nonumber\\
e^\phi&=& g_s z^{-1} = \frac{g_s}{\sin\theta'} \cos\alpha\ ,\ \
C=0 \nonumber\\
e^{2A}&=& \frac{\cot\theta'}{\tan\alpha}\ ,\label{BCinterp}\eea
and $h$ is the same as in (\ref{flow}).  These results are in fact identical
to those of \cite{Frey:2003sd}, which directly considered supersymmetry 
parameters with varying $\alpha$ and $\beta=\pi/2$.  Note, however, that
here we have seen that vanishing asymptotic $C$ requires $C$ to vanish
everywhere, which seems to restrict the varying $C$ solutions of 
\cite{Frey:2003sd}.

We see, therefore, how limits which turn off some variation 
of the supersymmetry parameter are singular.  In fact, this should be 
precisely true for the restriction of an $SU(2)$ structure to an $SU(3)$
structure.  This is one reason that it is useful to have a direct analysis
of the $SU(3)$ structure case.

\section{$(p,q)$5/D3 Bound State}\label{s:bound}

Now we will use the integrated scalars (\ref{flow}) above to derive the
supergravity solution corresponding to a bound state of $(p,q)$5-branes and
D3-branes.  This solution has been obtained previously
by a clever use of zero modes in the equations of motion 
\cite{Cederwall:1999zf} and
by application of dualities to known brane solutions 
\cite{Mitra:2000wr,Mitra:2001qh}.  The equivalence of these solutions was
demonstated in \cite{Gran:2001ga}.  After we have determined the solution 
explicitly, we can compare to the asymptotically flat bound state solution of 
\cite{Mitra:2000wr,Mitra:2001qh}.  

We start by specifying the six dimensional metric.  These six dimensions 
separate into those parallel and transverse to the $(p,q)$5-brane.  Since
we are working with the flat space solutions, the orthogonal directions should
have an $SO(4)$ symmetry for a single-center solution.  
Similarly, the two directions parallel to the
5-brane but orthogonal to the D3-branes (ie, the directions of the 5-brane
worldvolume field strength) should be translationally invariant.  Restricting
$a,b=4,5$ along the 5-brane and $m,n=6-9$ orthogonal, the metric becomes
\be\label{6dmetric}
ds_6^2=e^{2B_1}\t g_{ab}dx^a dx^b + e^{2B_2}\delta_{mn}dx^m dx^n\ .\ee
Then all of the supergravity scalars, $\alpha$, $\beta$, and the warp 
factors $A$, $B_{1,2}$ depend only on the radial direction $r^2=\delta_{mn}
x^m x^n$.  The complex structure can be calculated by wedging the vielbein
$J=e_{\hat 4}\wedge e_{\hat 5}+e_{\hat 6}\wedge e_{\hat 7}+e_{\hat 8}
\wedge e_{\hat 9}$.  (Hats denote tangent space indices.)

Because the 3-forms $H,F$ have vanishing Bianchis (up to sources
at the origin), their integral over $S^3$s concentric with the origin is
independent of radius; we normalize as
\be\label{bound3form1}
\oint_{S^3} H = 2p\Omega_{S^3}\ ,\ \ \oint_{S^3} F = 2q\Omega_{S^3}\ ,\ee
where $\Omega_{S^3}$ is the volume of a unit $S^3$\footnote{This normalization
absorbs the 5-brane charge into $p,q$.}. The allowed components
are therefore
\bea
H_{mnp} = e^{-2B_2} p \epsilon_{mnp}{}^q\del_q g&,&
F_{mnp} = e^{-2B_2} q \epsilon_{mnp}{}^q\del_q g\ , \ \ 
g=1/r^2\nonumber\\
H_{r45}= e^{-2B_1}\epsilon_{45}\del_r g_1 &,&
F_{r45}= e^{-2B_1}\epsilon_{45}\del_r g_2\ .\label{bound3form2}\eea

Consider (\ref{HFbeta}).  The $m=r$ component
along with the scalar relations (\ref{flow}) immediately give
(dropping the integration constants by gauge choice)
\be\label{g1g2}
g_1=-g_s\tan\theta g_2\ .\ee  Then the same component of (\ref{HFalpha}) and
(\ref{flow}) can be integrated to give
\be\label{g1g2int}
g_1= \frac{\sin\theta}{\sin(2\theta')}\cos(2\alpha)\ ,\ \ 
g_2= -\frac{\cos\theta}{g_s\sin(2\theta')}\cos(2\alpha)\ .\ee

We can continue by considering (\ref{dJ}).  The component along the three
angular directions vanishes on both sides, $dJ$ by computation, and the
right hand side by (\ref{flow}) and (\ref{g1g2}).  The component along
the radius and two of the angles yields a differential equation for the
$B_2$ warp factor; assuming asymptotic flatness gives
\be\label{B2}
e^{2B_2}= \frac{\sin^2\theta}{\cos\theta'\cos\theta}
\frac{\sin\alpha\cos\beta}{\sin^2\beta}\ .\ee
The $r45$ component gives a differential equation for $B_1$; however, this
equation exhibits the radius explicitly due to the appearance of the 
fluxes.  To solve it, we need to know the radial dependence of the angle
$\beta$ (and therefore all the other scalars).

This we find from the angular components of (\ref{HFalpha},\ref{HFbeta}).
Because $\alpha$ depends only on $r$, we find immediately from (\ref{HFalpha})
that $\tan\theta = qg_s/p$. Then we can integrate (\ref{HFbeta}) to find
\be\label{betar}
\sin\beta = \frac{r\sin\theta}{[r^2+p\cos\theta\cos\theta']^{1/2}}\ .\ee
We can now integrate the equation for $B_1$; coincidentally, the explicit
$r$ dependence now can be reabsorbed into $\beta$, so
\be\label{B1}
e^{2B_1} = \frac{\cos\theta}{\cos\theta'}\frac{\sin\alpha}{\cos\beta}\ .\ee
We can now write the radial dependence of all the scalars, warp factors,
and fluxes.

The final thing to determine is the asymptotic angle $\theta'$.  As discussed
in section \ref{ss:other}, we should also have to apply the 5-form Bianchi
identity to solve the equations of motion.  As a substitute, however,
we will note that we want a brane bound state with $n$ units of D3-brane
charge (using the same normalization as in equation (\ref{bound3form1})).
Therefore, asymptotically, where $H\wedge F$ vanishes, we should have
\be\label{5formquant}
\oint_{T^2\times S^3} \t F_5 = 2n\Omega_{S^3} V_{T^2}\ ,\ee
where $V_{T^2}$ is the volume of the torus wrapped by the 
5-branes\footnote{This seems to be an odd quantization condition; however,
we can derive it by considering the eight dimensional theory reduced along
the torus.  It also happens to give $n$ the same dimensionality as $p,q$.}.
Using the self-duality of the 5-form, this condition can be written as 
\be\label{5formbc}
e^{2B_1}e^{2B_2}e^{-4A}\del_r h = \frac{2n}{r^3}\ ,\ee
Comparing to the radial behavior of the scalars (\ref{betar}) then yields
$\tan\theta'=ng_s/\sqrt{p^2+q^2g_s^2}$.

We can also check that the remaining constraints are satisfied.  Indeed,
(\ref{Hsingle},\ref{Fsingle}) are just linear combinations of 
(\ref{HFalpha},\ref{HFbeta}) if we use (\ref{flow}).  We will not explicitly
discuss (\ref{12project},\ref{dstarJ},\ref{domega}), but they are 
straightforward, though slightly tedious, to check.

Finally, we note that we can compare our solution to that obtained by 
dualities in \cite{Mitra:2000wr,Mitra:2001qh} (with vanishing asymptotic
RR scalar).  That solution is written in terms of three warp factors,
$H,H',H''$ in the notation of \cite{Mitra:2000wr,Mitra:2001qh}.  By
comparing the radial dependence, it is easy to see that
\be\label{Hwarps}
H=\tan\theta'\tan\alpha\ ,\ H'=\tan^2\theta\cot^2\beta=
\frac{\sin^2\theta'}{\cos^2\alpha}\ ,\ H''=\frac{\sin^2\theta}{\sin^2\beta}\ .
\ee
Then the solutions match exactly (up to sign conventions).  
While this bound state solution has been obtained before, both from dualities
and the equations of motion, it is worth noting that our methods may be more
easily generalized to geometries with less symmetry and also to boundary
conditions more appropriate for gauge-string theory dualities.

\section{Summary}\label{s:summary}

In this paper, we have presented constraints on solutions of type IIB
supergravity with four dimensional Poincar\'e invariance and $\N=1$
supersymmetry based on an $SU(3)$ structure.  The $SU(3)$ structure 
is a limit of $SU(2)$ structures which have been recently presented in
\cite{Dall'Agata:2004dk}; however, the limit is degenerate, so a direct 
analysis is desirable.  Additionally, $SU(3)$ structures have nongeneric
behavior among $SU(2)$ structures; for example, all $SU(3)$ structures have
an integrable complex structure.  

In section \ref{s:radial}, we presented a solution to the relations among
the scalars in the case that all the scalars depend only on a single (radial)
direction.  It is our hope that this solution, with appropriately modified
boundary conditions, will be useful in constructing supergravity duals to
renormalization group flows.  We have used these results to derive the
supergravity solution for the $(p,q)$5/D3-brane bound state.

\acknowledgments
I would like to thank M. Caldarelli, G. Dall'Agata, J. Gauntlett, J. Gomis,
M. Gra\~na, and M. Schulz for useful communications.  This work was supported
by the John A. McCone Fellowship in Theoretical Physics at the California
Institute of Technology.

\appendix
\section{Conventions and Identities}\label{a:conventions}

In this paper, we work in string frame with usual string conventions for
the field strengths (see \cite{Hassan:1999bv}).  Note that the gravitino,
dilatino, and supersymmetry parameter are redefined in transforming from the
Einstein frame.  Indices $M,N$ are for the full 10 dimensions,  $\mu,\nu$ for
the 4 Poincar\'e invariant dimensions, and $m,n$ for the internal dimensions.
Hats denote tangent space indices.  We work in a signature in which timelike
norms are negative.

Our differential form conventions are as follows:
\bea
\epsilon_{012\cdots (d-1)} &=& +\sqrt{-g}\textnormal{ for $d$ dimensions}
\nonumber\\
T_{[M_1\cdots M_p]} &=& \frac{1}{p!} \left( T_{M_1\cdots M_p} \pm
\textnormal{permutations}\right)\nonumber\\
\left(\star T\right)_{M_1\cdots M_{d-p}}&=& \frac{1}{p!}\epsilon_{M_1\cdots
M_{d-p}}{}^{N_1\cdots N_p}T_{N_1\cdots N_p}\nonumber\\
T&=& \frac{1}{p!} T_{M_1\cdots M_p} dx^{M_1}\cdots dx^{M_p}\ .\label{forms}
\eea
Wedges and exterior derivatives are defined consistently with those 
conventions.  We choose $\star\t F_5=\t F_5$ for the self-duality of the 
5-form.

Gamma matrices in tangent space have the algebra 
$\{\Gamma^{\hat M},\Gamma^{\hat N}\}=2\eta^{\hat M\hat N}$.  With these
conventions, a Majorana basis is real and symmetric for spacelike indices
and antisymmetric for time.  Gammas can be converted to coordinate indices
with the vielbein.  We define $\Gamma^{M_1\cdots M_p}=\Gamma^{[M_1}\cdots
\Gamma^{M_p]}$.
The chirality is given by
\be\label{10dchiral}
\Gamma_{(\widehat{10})}= \Gamma^{\hat 0}\cdots\Gamma^{\hat 9}=
\frac{1}{10!}\epsilon_{M_1\cdots M_{10}}\Gamma^{M_1\cdots M_{10}}\ .\ee
The gravitino and supersymmetry parameter have positive chirality; this
choice agrees with the 5-form self-duality above.

We can decompose the $\Gamma$ matrices as
\be\label{decomp}
\Gamma^\mu = \gamma^\mu\otimes 1\ ,\ \ \Gamma^m = \gamma_{(\hat 4)}\otimes
\gamma^m\ee
with 4 and 6 dimensional chirality $\gamma_{(\hat 4)}=-i\gamma^{\hat 0}\cdots
\gamma^{\hat 3}$, $\gamma_{(\hat 6)}= i \gamma^{\hat 4}\cdots\gamma^{\hat 9}$.
The $\gamma^\mu$ have the same symmetry and reality properties as $\Gamma^M$,
while the $\gamma^m$ are imaginary and antisymmetric.

A number of $\gamma$ identities are very useful.  A comprehensive list
of (anti)commutators appears in \cite{Candelas:1984yd}, although there is
at least one typographical error.  It is necessary to replace
\be\label{typo}
[\gamma_{mnp},\gamma^{rst} ]= 2\gamma_{mnp}{}^{rst}-36\delta_{[mn}^{[rs}
\gamma_{p]}{}^{t]}\ .\ee
Additionally,
\be
\gamma_{mnpq} = \frac{i}{2}\gamma_{(\hat 6)} \gamma^{rs}\epsilon_{mnpqrs}\ ,\
\ \chi^\dagger\gamma_{mnpqrs}\chi = -i\epsilon_{mnpqrs}\ee
for positive chirality $\chi$.  

The Fierz identities that we use come from expanding in terms of the complete
set of $\gamma$ matrices.  Specifically, we find
\be\label{fierz}
\chi\chi^\dagger = \frac{1}{8}-\frac{i}{16}J_{mn}\gamma^{mn}
-\frac{i}{16}J_{mn}\gamma^{mn}\gamma_{(\hat 6)} +\frac{1}{8}\gamma_{(\hat 6)}
\ee
for the normalized positive chirality spinor $\chi$ used in the text.
This identity can be used to show that $J_m{}^nJ_n{}^p=-\delta_m^p$ and also
that
\be\label{fierz1}
6J_{[mn}J_{pq]} = \epsilon_{mnpq}{}^{rs}J_{rs}\ .\ee

\bibliographystyle{h-physrev4}\bibliography{su3structure}
\end{document}